\documentclass[aps,prb,twocolumn,groupedaddress]{revtex4}
\usepackage{subfigure}
\usepackage{amsmath}
\usepackage{graphicx}
\usepackage{rotating}
\usepackage{amsfonts}
\usepackage{amssymb}
\begin{document}
\title{Effect of doping and pressure on magnetism and lattice structure of 
Fe-based superconductors}
\author{M.D. Johannes, I.I. Mazin, D.S. Parker}
\affiliation{Code 6393, Naval Research Laboratory, Washington, D.C. 20375}
\date{Printed on \today}

\begin{abstract} Using first principles calculations, we analyze structural and magnetic trends as a function of 
charge doping and pressure in BaFe$_2$As$_2$, and compare to experimentally established facts.  We find that 
density functional theory, while accurately reproducing the structural and magnetic ordering at ambient pressure, 
fails to reproduce some structural trends as pressure is increased. Most notably, the Fe-As bondlength which is a 
gauge of the magnitude of the magnetic moment, $\mu$, is rigid in experiment, but soft in calculation, 
indicating residual local Coulomb interactions.  By calculating the magnitude of the magnetic ordering energy, we 
show that the disruption of magnetic order as a function of pressure or doping can be qualitatively reproduced, 
but that in calculation, it is achieved through diminishment of $|\mu|$, and therefore likely does not reflect the 
same physics as detected in experiment.  We also find that the strength of the stripe order as a function of 
doping is strongly site-dependent: magnetism decreases monotonically with the number of electrons doped at the Fe 
site, but increases monotonically with the number of electrons doped at the Ba site.  Intra-planar magnetic 
ordering energy (the difference between checkerboard and stripe orderings) and interplanar coupling both follow a 
similar trend.  We also investigate the evolution of the orthorhombic distortion, $e=(a-b)/(a+b),$ as a function 
of $\mu$, and find that in the regime where experiment finds a linear relationship, our calculations are 
impossible to converge, indicating that in density functional theory, the transition is first order, signalling 
anomalously large higher order terms in the Landau functional.

\end{abstract}

\maketitle

\section{Introduction}

The magnetic properties of the Fe-based superconductors are believed to be the key
to understanding their normal and superconducting properties\cite{MS}. Yet a
consensus about the microscopic physics of magnetism in these materials is
still lacking. There are several widely held ideas that are arguably
supported by most researchers in the field. First, the magnetism is intimately
related to the crystal structure, both in terms of the Fe-As bond length,
which is reduced when the local magnetic moment on Fe disappears (a simple
reflection of the magnetostrictive nature of Fe),  and in terms of an
orthorhombic distortion in the magnetically ordered state (it is nearly
universally believed that the distortion is driven by magnetism and not the
other way around). The orthorombicity of up to 1\% is comparable with, say,
the rhombohedral distortion of 1.2\% in FeO upon the antiferromagnetic ordering.

Second, although initial opinions about the origin of the magnetic ordering
in Fe pnictides stretched from a spin-Peierls philosophy
\cite{Firstpaper,Chubukov} to Mott physics\cite{Si}, it has now been
recognized that while the local magnetic moments on Fe are formed
independently of the fermiology, their mutual interaction is largely
controlled by the itinerant electrons' response and by the Fermi surface
geometry\cite{FeTe,optics}. A corollary of this fact is that when the
long-range order is destroyed (whereupon superconductivity usually emerges), the
system should be described as \textit{para}magnetic, a collection of
disordered magnetic moments, rather than \textit{non}magnetic, with the
magnetic moment uniformly suppressed, as in non-spin-polarized density
functional calculations. Particularly questionable are attempts to describe
the evolution of magnetic (and therefore crystallographic) properties when
magnetism is suppressed (for instance, by pressure). It has been established
\cite{Yildirim,Yin,PRB,Jesche} that density functional theory within the generalized gradient
approximation (DFT-GGA) describes the crystal structure (as
well as the phonon spectra) of the parent compounds very accurately at ambient pressure, as long as 
full magnetization is allowed. It is not clear, however, whether DFT-GGA will
work as well under pressure (the argument above suggests it may not)
One purpose of this paper is to address this question.

Another unresolved and important question is the underlying mechanism by which 
the AFM order is destroyed by external means. Experimentally, one can proceed in 
three different ways. Chronologically the first method used was formally similar 
to that used in superconducting cuprates: charge doping. Naturally, it was 
implicitly assumed that, as in cuprates, charge doping increases the number of 
carriers, improves the metallic screening and renders the system less strongly 
interacting, and thus, less magnetic. In accordance with this concept, it was 
discovered \cite{canfield} that Ni (which donates two electrons) is about twice 
more efficient in destroying the long-range magnetism as Co (which donates only 
one), and that electron doping (substituting O by F, or Fe by Co and Ni) has 
qualitatively the same effect as hole doping (substituting Ba by K). However, 
later it was found that pressure and/or strain can lead to essentially the 
same effect \cite{alireza,Kimber}, suggesting that the carrier concentration is 
not the only, and maybe not even the most important change brought about by the 
chemical doping.  This view was further reinforced by the fact that partial 
substitution of As by P (which exerts chemical pressure on Fe) has again the 
same effect \cite{ren}. Finally, it was also shown that diluting the Fe plane by 
nonmagnetic atoms, such as Ru, again destroys the magnetic order and triggers 
superconductivity \cite{bharathi,mcguire}.

DFT calculations can account for the last two effects, at least
on the qualitative level: both physical (volume reduction) or chemical
(reducing the iron-pnictogen height) pressure in calculations reduces the
tendency to magnetism. However, it is not immediately clear what effect charge
doping should have on magnetism inside DFT.  In particular, if the mechanism of 
suppression is not the same as in cuprates, why would both hole and
electron doping have the same, negative effect on magnetism? Answering this
question is the second goal of this paper. We find that DFT does show the same
qualitative behavior, doping electrons at the Fe site in BaFe$_{2}$As$_{2}$
depresses the magnetism, as does doping holes at the Ba site, while,
intriguingly, doping holes on the Fe site and electrons on the Ba site
enhances it.

Last but not least, there have been experimental indications of a $linear$ as 
opposed to quadratic relation between the orthorhombic order parameter, 
$e=(a-b)/(a+b),$ and the measured magnetic moment. Such a relationship is 
formally prohibited by symmetry in the Landau theory, being only possible if the 
neutron measured magnetic moment is not the actual order parameter, or if the 
Landau functional includes anomalously large higher order terms and thus the 
quadratic regime extends only over very small magnetization. This anomalous 
behavior has been observed both as a function of temperature \cite{Geibel} and 
as a function of doping \cite{DaiCe}. While there is no guarantee (as discussed 
above) that DFT is capable of describing this magnetic phase transition 
correctly, it is still of interest to see whether the Landau functional 
\textit{as calculated in DFT} does have anomalously large high-order terms. This 
is the third issue we address in this publication.

\section{Methods} All calculations as a function of pressure were carried out using the Vienna ab-initio 
simulation package (VASP) \cite{vasp}, a projector augmented wave (PAW) based pseudopotential formalism.  We 
employed the generalized gradient approximation (GGA)to the exchange potential of Perdew-Burke-Ernzerhof 
(PBE)\cite{pbe}.  We fully relaxed a series of structures (both lattice and internal coordinates) at a variety of 
volumes and extracted the pressure by fitting to an equation of state. All calculations as a function of doping 
were carried out using Wien2k \cite{Wien2k}, which employs an APW+lo (augmented plane wave plus local orbitals) 
basis set, again using PBE-GGA .  The lattice coordinates used were $a$ = 5.576, $b$= 5.616, $c$ = 12.950 \AA, 
and $z_{As}$ = 0.8972 (as a fraction of $c$), corresponding to the fully relaxed structure at zero pressure 
described previously.  To simulate charge doping without using a supercell, we employed the virtual crystal 
approximation (vca). This technique involves replacing each atom of a certain type in the unit cell with a 
fictitious element with a non-integer atomic number.  For electron doping at the Fe site, we replace $Z=26$ with 
$Z=26+x$ (toward Co) and for hole doping at the Ba site we use a element with $Z=56-x$ (toward Cs), using the same 
crystal structure.  The number of electrons in the systems is increased commensurately, so that overall charge 
balance is maintained (alternate hole/electron doping at the Co/Ba site is achieved by simply subtracting/adding 
to $Z$).  For calculations of intra- and interplanar coupling, we used two separate symmetries, Cmmm (space group 
66) for the observed antiferromagnetically stacked stripe ordering, Cmma (space group 67) for the 
ferromagnetically stacked stripe order, and I$\overline{4}$m2 (space group 119) for the checkerboard ordering.

\section{Structure as a function of pressure}

\begin{figure}[ptb]
\includegraphics[width = 0.85\linewidth, angle=270]{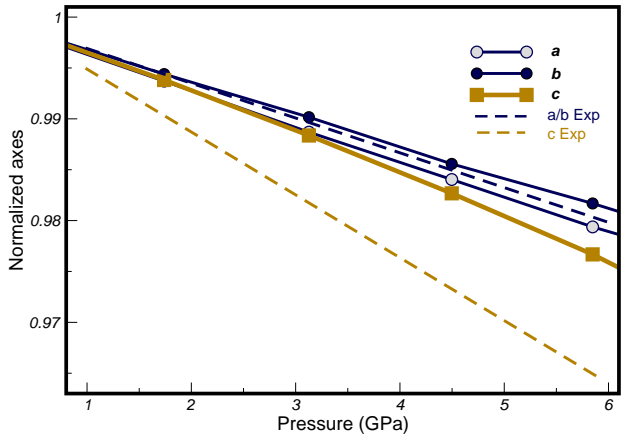}\caption{A comparison of the
lattice parameters $a$ (short in-plane axis), $b$ (long in-plane axis) and $c$
(out-of-plane axis) to experiment. All parameters are normalized to the zero
pressure value; the experimental lines are taken from Ref. \cite{Kimber}.}
\label{axes}
\end{figure}

In Ref. \onlinecite{Kimber}, Kimber \textit{et al.} found that both doping and
pressure cause the lattice parameters to decrease linearly. The Fe-As bond is
found to be extremely rigid, in good agreement with a previous experimental
study \cite{RotterAC}, while the As-Fe-As angle shrinks substantially with
increasing pressure. Additionally, they find no indication of a structural
anomaly or even change in structural trends occurring around the critical
pressure or critical doping. Our DFT calculations of $a$, the parameter
connecting FM-aligned spins, and $b$, the parameter connecting AFM-aligned
spins, show very good agreement with the single in-plane lattice parameter
measured by experiment (note that, due to overestimation of the static
magnetic moment, DFT maintains the magnetically induced orthorhombic
distortion up to $\sim$ 12GPa, whereas no long range ordering is
experimentally detected after 1.3 GPa). The agreement extends, both in
absolute value (not shown) and in trend with pressure (Fig.
\ref{axes}), throughout the measured range of pressures (1-6 GPa). The c-axis
parameter agrees well with experiment at zero pressure, but is stiffer in our
calculations than in experiment.

\begin{figure}
\includegraphics[width = 0.85\linewidth, angle=270]{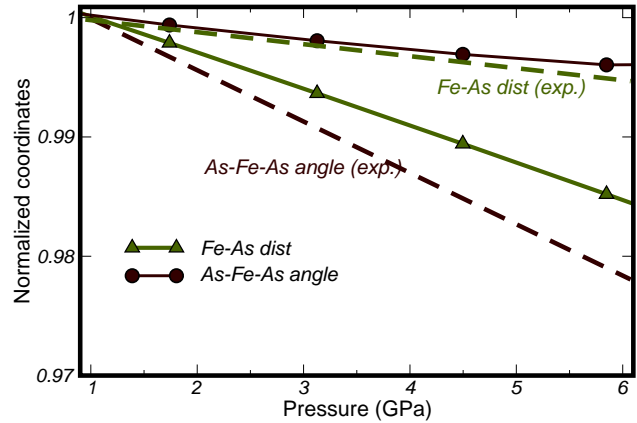}
\caption{The Fe-As bondlength and As-Fe-As angle normalized to their values at 1 GPa (solid lines) in comparison to 
similarly normalized experimental values from Ref. \onlinecite{Kimber} (dashed lines). In experiment, the Fe-As distance is nearly constant, while the As-Fe-As angle changes, whereas in DFT calculations, the opposite is true.}
\label{bonds}
\end{figure}

The main disagreement occurs in the Fe-As bond length and As-Fe-As angle (Figure \ref{bonds}). The
former shrinks linearly with pressure in DFT calculations, instead of
maintaining the observed constant value. The As-Fe-As angle, on the other
hand, is rather constant over the pressure range, whereas in experiment it
decreases. Both discrepancies are due to a single factor: the perpendicular
height of the As atom above the Fe plane (the in-plane component of the Fe-As
bondlength is determined by $a$ and $b$ which both match well with
experiment). This height scales linearly with the magnetic moment of the Fe
atom, $\mu$. The physical meaning of this is clear: as discussed, the Fe ion
is characterized by a large magnetostrictive effect; compressing the ion
results in a loss of the local magnetic moment. The Fe-As bond length controls
the chemical pressure on Fe and thus is strongly correlated with the moment.
The constant bondlength in experiment reveals that the magnitude of the
magnetic moment does {\it not} change under pressure, indicating that the
suppression of magnetic ordering occurs through increased spin fluctuations
and orientational disorder rather than through an actual decrease in the
absolute magnitude of the moment. DFT does not capture this effect,
compensating instead by decreasing the overall moment. The calculated As-Fe-As angle
suffers similarly from a decrease in As height that offsets the decrease in
$a$,$b$, leaving a relatively constant value. 

In view of the fact that the calculated equilibrium moment is larger than the 
experimentally measured one, one might assume that it would be $more$ rigid than 
in experiment. The fact that the opposite relationship takes place tells us that 
while DFT overestimates the ordered moment, it underestimates the local moment. 
In retrospect, this is not that surprising because there exist residual Coulomb 
correlations in the system (DMFT calculations in the 1111 systems \cite{georges} 
indicate about 70\% mass renormalization due to local Coulomb correlation, a 
small but not negligible number, which enhances the tendency toward local magnetism 
\cite{petukhov}.

\section{Magnetism as a function of pressure}

\begin{figure}[ptb]
\includegraphics[width = 0.86\linewidth, angle=270]{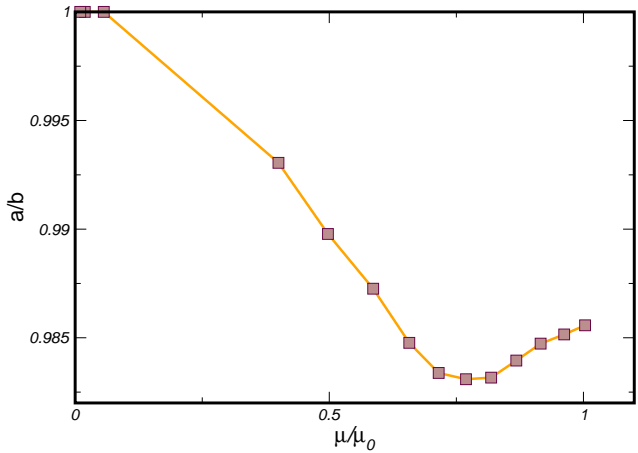}\caption{The ratio
of the $a$(short) to $b$ (long) lattice parameters as a function of the
relative magnetic moment $\mu/\mu_{0}$, where $\mu_{0}$ is the magnetic moment
at zero pressure.  No calculations could be converged between $\mu$=0 and $\mu$=0.4, 
indicating a first order transition.}
\label{abmu}%
\end{figure}

Outside the pressure range explored by Ref. \onlinecite{Kimber}, we find that
the $a$ and $b$ lattice parameters decrease in a non-linear fashion. The short
Fe-Fe (FM) bond does not decrease monotonically, while the long (AFM) bond
does. This causes the ratio of the two to reach a definite minimum and is the
origin of the minimum in $e=(a-b)/(a+b)$ around 6 GPa, corresponding to 1.55
$\mu_{B}$ in Fig. \ref{abmu}. (The qualitative behavior of $a$ and $b$ as a
function of pressure matches very well with a previous DFT study
\cite{nakamura} of the 1111 compound, LaFeAsO). The calculated minimum in $e$
is a consequence of moving from a magnetic state to a non-magnetic state via
full suppression of the magnetic moment. As discussed previously, the magnetic
moment is rigid in experiment, so the calculated behavior is not expected to
manifest in real systems. The original intent of performing this calculation
was to examine $e$ in the range where it increases linearly with $\mu$, as
seen in experiment \cite{DaiCe,Geibel}, $i.e.$ as $\mu$ grows away from zero.
Unfortunately, we find that calculations in this region are essentially
impossible to converge, suggesting a Landau functional with anomalously large
high-power terms. In other words, the AFM phase transition induced by pressure
or chemical pressure in the calculations seems to be first order, at least at
the level of DFT-GGA.

\begin{figure}[ptb]
\includegraphics[width = 0.85\linewidth, angle=270]{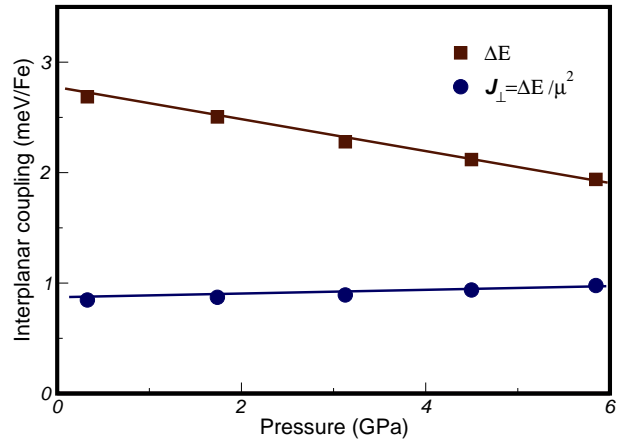}\caption{Interplanar
coupling, calculated as the difference in total energies between a system in
which the in-plane stripe order is anti-aligned in successive planes and a
system in which in-plane stripe orders are aligned in successive planes.}
\label{pressJ}
\end{figure}

We also investigated the interplanar coupling (the total energy difference between stripe layers stacked 
antiferromagnetically and ferromagnetically) as a function of pressure. If the coupling were a result of 
superexchange between Fe layers (whether directly through As-As hopping or through Ba atoms), one would expect it 
to increase as the layers are pushed closer together. As can be seen in Fig. \ref{pressJ}, there is a very slight 
increase in $J_{\perp}$, defined as $\Delta E=$ $J_{\perp}\mu^{2},$ as the pressure increases, but it is offset by 
a decrease in the magnetic moment, leaving the net coupling parameter essentially constant ( even {\it decreasing} 
very slightly) across the pressure range of 0-6 GPa. In conjunction with the fact that we find that energy difference 
between the checkerboard and stripe in-plane magnetic configurations decreases with pressure (not shown), these results are 
again consistent with a picture in which increased spin fluctuations destroy the long range order. However, as 
pointed out earlier, the decrease in the magnitude of $\mu$ as calculated by DFT may not accurately represent 
reality. It seems more likely that $|\mu|$ is constant, but increasingly fluctuates with pressure. In this case, 
the interplanar coupling would indeed increase with pressure and the observed suppression of magnetic long-range 
order must have a different origin, perhaps stemming from in-plane fluctuations.

\section{Magnetism as a function of doping}

\begin{figure}[ptb]
\includegraphics[width = 0.95\linewidth]{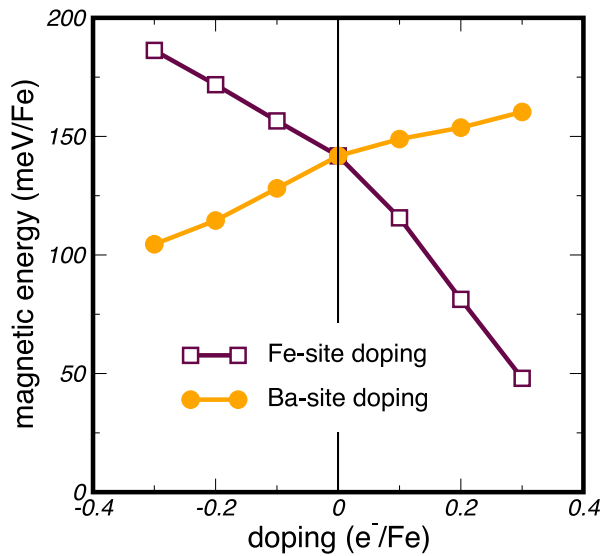}\caption{The magnetic
energy, defined as the total energy of the system in the magnetic stripe phase
minus the total energy of the nonmagnetic system, as a function of hole and
electron doping. Doping at the Fe site and on the Ba site for
are shown for the virtual crystal approximation. Filled symbols show supercell
calculations.}
\label{vca}
\end{figure}

One way to gauge the strength of the tendency toward magnetism is to evaluate the energy difference between a magnetic and 
a nonmagnetic (no local moments) solution. We have calculated this energy difference (Fig. \ref{vca}) by using the virtual 
crystal approximation imitating the Co doping on the Fe site and the K doping on the Ba site (see Methods section for 
details) using the relaxed structure at ambient pressue, $i.e.$ the effects of charge doping on the structure were not 
accounted for. We have further verified (Fig. \ref{vca}) that supercell calculations for Ba$_{2}$Fe$_{3}$CoAs$_{2}$ are 
quantitatively consistent with the VCA, and for BaKFe$_{4} $As$_{4}$ semi-quantitatively consistent.

Our results show that, in agreement with the experiment, both types of doping weaken the magnetism (reduce the 
magnetization energy). But, we also found that extending our VCA calculations onto the opposite sides of the phase 
diagrams, that is, introducing holes on Fe sites or electrons on Ba sites, the trend simply continues, so that in 
these two case the magnetism is $enhanced.$ This same trend was found for a DFT study of the Sr-based 122 compound 
\cite{deepa,kim}.  Neither of the regimes precisely corresponding to our calculations has been accessible so far 
experimentally. Hole doping on the Fe site formally corresponds to Mn or Cr substitution. These indeed strengthen 
the magnetism (in agreement with our prediction) \cite{deepa,kim}, but these dopants are likely themselves to have 
large local magnetic moments as impurities, and it is fairly possible that this is the reason for the experimental 
behavior, and not charge doping $per$ $se.$ Substituting Ba (or even better, Sr) by a rare earth like La or Yb 
seems to be chemically natural (\textit{cf.} superconducting cuprates or colossal magnetoresistance manganites), 
yet so far there has been no success in achieving it. The DFT prediction is that such doping will enhance or at 
least not suppress the magnetism.  It should be noted that the increase/decrease in magnetic energy occurs in 
conjunction with, and obviously partially due to, an increase/decrease in $|\mu|$.

Apart from local magnetism, the actual long range order depends on exchange 
interactions.  These fall into two categories, the in-plane interactions (which, 
in these systems, appear to be long range \cite{antropov,yildirim2} and 
non-Heisenberg \cite{antropov,yaresko}), and the interplanar coupling.

\begin{figure}[ptb]
\includegraphics[width = 0.95\linewidth, angle=270]{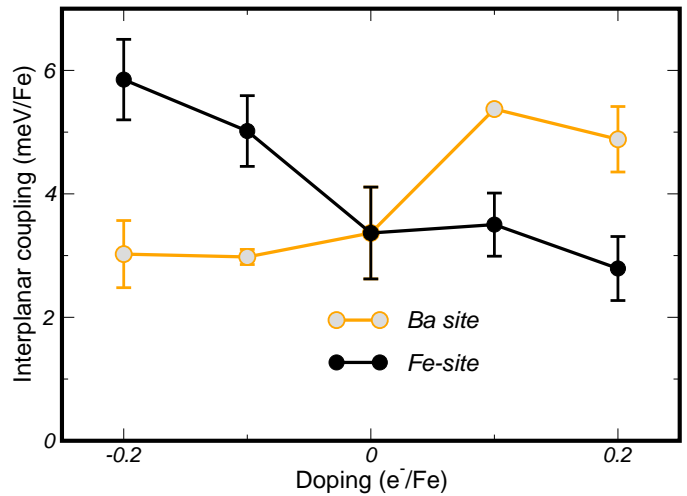}\caption{Interplanar
coupling, calculated as in Fig.\ref{pressJ}, but as a function of doping
rather than pressure. The virtual crystal approximation is used, with hole
doping taking place at the Ba site and electron doping taking place at the Fe
site}
\label{dopeJ}
\end{figure}

In particular, it has been suggested that the increase in spin fluctuations is due to an increased 
two-dimensionality brought about by a decrease in interplanar coupling \cite{harriger}. We have calculated this 
coupling for both hole and electron doping, again using the virtual crystal approximation.  The interplanar 
coupling does vaguely decrease in both directions (see Fig.\ref{dopeJ}) with the conventional doping sites (holes 
on Ba, electrons on Fe), although within the error bars, set by total energy convergence in our calculation, the 
trend can be considered as entirely flat.  The trends show the same site-dependence as the previously calculated 
magnetic energies. This is consistent with, albeit not proof of, the contention that magnetism is affected by the 
degree of two-dimensionality.  

As mentioned, the intraplane interactions are long-range and non-Heisenberg, therefore instead of mapping them 
onto simplified models like Heisenberg or Ising, we look directly at the magnetic ordering energy (energy 
difference between magnetic and non-magnetic) states.  We find that the intraplane coupling strongly decreases as 
a function of doping in the conventional scheme and yet again shows a strong site dependence.  At zero doping, the 
scale of the intra-planar coupling is an order of magnitude greater than the interplanar coupling, but by $x 
\approx$ 1.5 (where $x$ is the number of electrons per Fe), the energy advantage of the stripe order over 
checkerboard has disappeared entirely.  Collectively, the doping calculations point to a picture in which the  
primary influence of adding or subtracting charge (as with pressure) is to increase spin fluctuations.

\begin{figure}
\includegraphics[width=0.85\linewidth, angle=270]{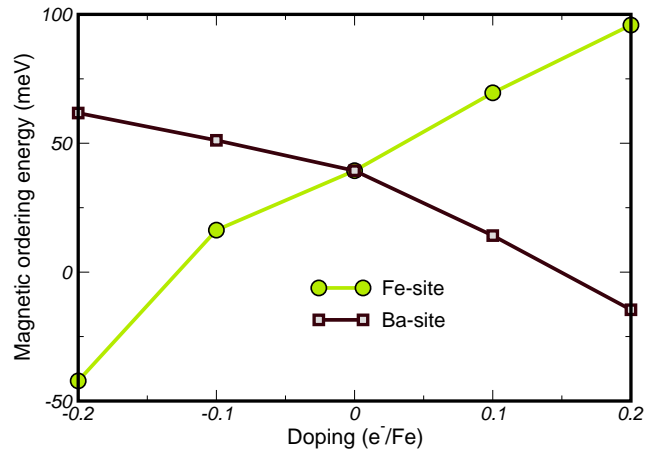}
\caption{The magnetic energy difference between in-plane stripe and in-plane checkerboard orderings as a function of 
doping in the virtual crystal approximation.}
\label{inplane}
\end{figure}

\section{Conclusions}

To summarize, we have extended the familiar DFT-GGA calculations to address several
issues not addressed previously. Our findings are as follows:

(1) Although spin-polarized DFT-GGA predicts the equilibrium crystal structure
at zero pressure exceedingly well, it becomes increasingly worse with
pressure. Specifically, the Fe-As bond is significantly softer in the calculations than in
the experiment. We interpret this as evidence that the local magnetic moment
(as opposed to the average ordered moment) is smaller in the calculations, not
larger, than in the experiment, and ascribe this to residual local Coulomb correlations.

(2) The antiferromagnetic interlayer coupling is mainly constant as a function of pressure in DFT,
which indicates that it is not of pure superexchange origin. We interpret this
as an indication of at least two competing interplanar interections, one
antiferromagnetic and one ferromagnetic (double exchange), whose pressure
dependencies cancel one another.

(3) The small magnetic moment regime in BaFe$_2$As$_2$ is inaccessible to DFT-GGA
calculations under hydrostatic pressure. This is likely an indication that
within DFT-GGA the phase transition is first order, and that the Landau
functional in the DFT-GGA has anomalously large high-order terms, consistent
with the fact that experimentally the scalar orthorhombic order parameter
follows the absolute value of the vector (magnetic) order parameter, and not
the square of the latter.

(4) The effect of doping strongly depends on the location of the doped charge. 
Electronic doping in the Fe plane or hole doping in the Ba plane reduces the tendency 
to form local moments, while hole doping in the Fe plane or electron doping in the Ba 
plane enhances it.  Although we are unaware of any Ba-plane electron doping 
experiments to date, we predict that this effect should be verifiable via an
increased magnetic ordering temperature and decreased superconductivity.

(5) The former two kinds of doping reduce the interlayer coupling, while the latter two enhance it.  The interplanar coupling is 
essentially insensitive to doping within conventional doping scheme (holes on the Ba site or electrons on the Fe site).

(6) Intraplanar interaction again shows strong site dependence, but decreases very strongly as a function of 
doping in either direction within the conventional doping scheme, further supporting the idea that the role of 
dopants in suppressing magnetism is to increase spin fluctuations.

We acknowledge funding from the Office of Naval Research.

\end{document}